\documentstyle[preprint,aps,epsf,floats]{revtex}

\begin{document}

\preprint{\tighten \vbox{\hbox{} }}

\title{On the Resummed Hadronic Spectra\\
       of Inclusive $B$ Decays}

\author{Adam K.\ Leibovich, Ian Low, and I.\ Z.\ Rothstein}

\address{
Department of Physics,
Carnegie Mellon University,
Pittsburgh, PA 15213}

\maketitle

{\tighten
\begin{abstract}

In this paper we investigate the hadronic mass spectra of inclusive
$B$ decays. Specifically, we study how an upper cut on the invariant
mass spectrum, which is necessary to extract $V_{ub}$, results in the
breakdown of the standard perturbative expansion due to the existence
of large infrared logs.  We first show how the decay rate factorizes
at the level of the double differential distribution. Then, we
present closed form expressions for the resummed cut rate for the
inclusive decays $B\rightarrow X_s\,\gamma$ and $B\rightarrow
X_u\,e\,\nu$ at next-to-leading order in the infrared logs. Using these
results, we determine the range of cuts for which resummation is
necessary, as well as the range for which the resummed expansion
itself breaks down. We also use our results to extract the leading and
next to leading infrared log contribution to the two loop differential
rate. We find that for the phenomenologically interesting cut values, 
there is only a small region where the calculation is
under control.  Furthermore, the size of this region is sensitive to
the parameter $\bar{\Lambda}$. We discuss the viability of extracting
$V_{ub}$ from the hadronic mass spectrum.
\end{abstract}
}


\newpage

\section{Introduction}

Inclusive $B$ decays are considered fertile ground for precision tests
of the standard model. The process $B \to X_u\,e\,\nu$ can be
used to extract the all important Cabibbo-Kobayashi-Maskawa (CKM) 
matrix element $V_{ub}$, while
$B\to X_s \gamma$ decays are important for discovering new
physics. However, the utility of experimental measurements of these
processes is bounded by our ability to control the theoretical
errors. Tremendous effort has gone into determining ways to calculate
these rates in a systematic fashion. Indeed, the algorithm for
calculating these rates is now part of the theoretical canon \cite{B}.
Unfortunately, experimental cuts complicate life for theorists. 
In particular, these cuts often force us to work near
the boundary of the phase space, where the aforementioned canonical
techniques break down.
Only now are we 
learning how to retool our calculations to accommodate
these highly non-trivial issues.

The complications arise if the cut forces us into a corner of phase
space, since the calculation can now depend on a new parameter, $\rho$,
which is a measure of the relative size of phase space of
interest. When this parameter becomes parametrically small, the
systematics of the calculation usually break down;
perturbative QCD corrections become enhanced by large logs of the form
$\log{\rho}$, while the non-perturbative expansion in $
\Lambda/m_b$ becomes an expansion in $\Lambda/(\rho\,m_b)$.

The most relevant cut rate arises  in semi-leptonic $B$ decays,
where one wishes to measure $V_{ub}$ by eliminating
the large background from charmed transitions.
To eliminate this background, we have a choice of variables with which
to cut. Perhaps the simplest choice is the
 electron energy, which  is  the oldest method
used for extracting $|V_{ub}|$. Unfortunately, as has been widely
discussed in the literature, such extractions are typically model
dependent since the rate in this window is sensitive to the 
Fermi motion of the heavy quark. There is no way to write down 
a meaningful theoretical error for such extractions.
It is only very recently that
a model independent method has been proposed, within a well defined
systematic scheme \cite{LLR}, that could lead to an extraction with
a well defined error. 

It is also possible to remove the background from 
charmed transitions by cutting on the hadronic invariant mass \cite{D}.
While this choice presents a greater experimental challenge, it 
benefits from the fact that, unlike the electron spectrum,
most of the $B\rightarrow X_u e \nu$ decays are expected to lie within
the region $s_H<M_D^2$. 
Furthermore, it is believed that 
even though both the invariant mass region  $s_H<M_D^2$
and electron energy regions $M_B/2>E_e>(M_B^2-M_D^2)/(2 M_B)$ receive
contributions from hadronic final states with invariant mass up 
to $M_D$, the cut mass spectrum will be less sensitive
to local duality violations.  This belief rests on the fact that  
 the contribution of
large mass states is kinematically suppressed 
for the  electron energy spectrum in the region of interest.

The goal of this paper is to study the  viability of extracting
$V_{ub}$ from the invariant mass spectrum.
In particular, we are interested in studying the breakdown
of the perturbative expansion for the cut rate, and whether
or not a reorganized expansion can be used reliably.

Building upon the work of Korchemsky and Sterman \cite{KS}, we begin
by discussing how the doubly differential decay rate factorizes in
moment space. We then use the recent results of \cite{LLR} to
calculate a closed form expression for the inverse Mellin transform at
next-to-leading logarithmic (NLL) order.  The result for the resummed
rate is presented in terms of the partonic as well as hadronic
invariant mass. This result is used to extract a piece of the two loop
rate, the size of which can be compared to the
Brodsky-Lepage-Mackenzie (BLM) two loop correction.  We then determine
the region of invariant mass where resummation is necessary as well as
the region where the reorganized expansion breaks down. We conclude
with a brief discussion of the phenomenology, saving a complete
discussion, including the effects of Fermi motion, for a later
publication\cite{prep}.

\section{Factorization in Hadronic Variables}

The problem of summing large threshold logarithms in perturbative
expansions, which arise due to incomplete KLN cancellation of the IR
sensitivity at the edge of phase space, has been addressed  
for various processes \cite{Srev}.  The technique relies on 
factorization, which allows for resummations via a renormalization
group equation. The factorization in $B$ decays has been 
previously discussed in leptonic variables in \cite{KS,AR}.  
Here we review the arguments that are germane to our discussion
of factorization in terms of hadronic variables. 

Consider the inclusive semi-leptonic decay of the $b$
quark into a lepton pair with momenta
$q=(p_e+p_\nu)$ and a hadronic jet of momenta $p_h$. It is convenient
to define the following {\em partonic} kinematic variables in the rest
frame of the $B$ meson $v=(1,\vec{0})$, 
\begin{equation}
\hat s_0=\frac{p_h^2}{m_b^2},\quad
h =\frac{2v\cdot p_h}{m_b},\quad x = \frac{2v\cdot p_e}{m_b},
\end{equation}
with phase space boundaries
\begin{equation}
0 \leq \hat s_0 \leq 1, \quad  2\sqrt{\hat s_0} \leq h \leq 1+\hat s_0, \quad
1-\frac{h}{2}-\frac12\sqrt{h^2-4\hat s_0}\leq x \leq
1-\frac{h}{2}+\frac12\sqrt{h^2-4\hat s_0}.
\end{equation}
In addition, it is customary to define the leptonic variables
\begin{equation}
y_0=\frac{2v\cdot q}{m_b},\quad
y=\frac{q^2}{m_b^2}.
\end{equation}
In terms of the leptonic variables, $\hat s_0=(1-y_0+y)$ and $h=2-y_0$, one
can see that in the endpoint region of the electron energy spectrum
when $x\to 1$ with $y<1$, the invariant mass of the jet  approaches
zero with its energy held fixed.  In addition, 
the jet hadronizes
at a much later time in the rest frame of the $B$ meson, due to the
time
dilation. Factorization exploits this  and separates the 
particular differential rate under consideration into subprocesses with
disparate scales. This factorization fails when the jet energy vanishes
in the dangerous region $y\to x\to 1$. However, this problematic region of 
phase space is suppressed because  the rate to 
produce soft massless fermions vanishes at tree level.

The infrared sensitive regions, which give rise to the large
logarithms, can be determined by constructing a {\em reduced} diagram,
as shown in Fig.~1.  According the Coleman-Norton theorem \cite{CN}, a
diagram at the infrared singular point must describe a physically
realizable process after contracting all off-shell lines to a
point. In the figure, $S$ denotes a soft blob which interacts with the
jet and the $b$ quark via soft lines. $J$ denotes the hadronic jet and
$H$ the hard scattering amplitude. Thus, the reduced diagram in Fig.~1
is simply a visualization of factorization. An important consequence
of this factorization is that the soft function $S$ is universal.
Moreover, it has been shown \cite{KS} that the soft function $S$
contains the non-perturbative structure function introduced in
Ref.~\cite{N}.  It is this universality that will eventually allow us
to eliminate the dependence on unknown non-perturbative hadronic
dynamics\cite{LLR,prep}.

\begin{figure}[t]
\centerline{\epsfysize=6truecm  \epsfbox{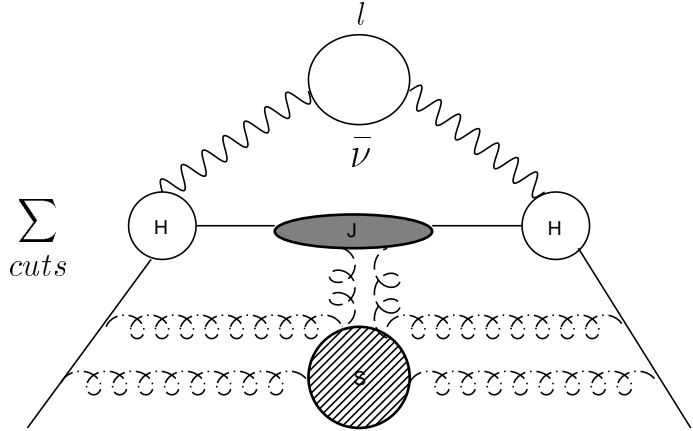} }
\tighten{
\caption[]{\it Reduced diagram for inclusive B decays.}}
\end{figure}

From the above discussions one can see that in terms of the variables
introduced earlier, factorization holds when $\hat s_0 \sim 0$ and $h
\sim 1$, or equivalently $1 \gg \hat s_0/h$. The typical momenta
flowing through the hard subprocesses are ${\cal O}(m_b)$. Thus, $H$
does not contain any large threshold logarithms and has a well-defined
perturbative expansion in $\alpha_s(m_b)$. The soft function $S$
contains typical momentum $k$, with $k^+\sim k^-\sim k_\perp = {\cal
O}(m_b\,\hat s_0/h)$.  By {\em soft} we mean soft compared to $m_b$,
but still larger than $\Lambda_{\rm QCD}$. For an energetic $u$ quark
moving in the '$-$' direction, the jet subprocess has typical momenta
$p$ such that $p^- \gg p^+, p_\perp$ with $p^+ ={\cal O}(m_b\,\hat
s_0/h)$, $p^2_\perp={\cal O}(m_b^2\,\hat s_0/h)$, and $p^-={\cal
O}(m_b)$.  In order to delineate between momentum regimes, a
factorization scale $\mu$ is introduced. The fact that the process is
independent of the factorization scale $\mu$ is utilized to sum the
large threshold logarithms in the soft and jet functions. The reduced
diagram for the inclusive radiative decays $B \to X_s\gamma$ is
exactly the same as above if we replace the lepton pair with a photon
and ignore the strange quark mass.

In terms of $\hat s_0$ and $h$, the triply differential rate, which
factorizes into hard, jet and soft subprocesses\cite{KS}, may be
written as
\begin{eqnarray}
&&\frac1{\Gamma_0}\frac{d^3\Gamma}{d\hat s_0\,dh\,dx} = \nonumber\\
&&\quad\quad 12\, (2 - h - x)(x + h - 1 - \hat s_0)
  \int_\xi^{M_B/m_b} dz\,S(z)\,m_b^2\,J[m_b^2\; h (z-\xi),\mu]\,
  H(m_b\,h/\mu), \\
&&\phantom{\frac{d^3\Gamma}{d\hat s_0\,dh\,dx}}
 \Gamma_0 = \frac{G_F^2}{192 \pi^3}|V_{ub}|^2 m_b^5,
\end{eqnarray}
where $\xi = 1 - \hat s_0/h$ is  analogous to the Bjorken scaling
variable in deep inelastic scattering.  $z=1+k_+/m_b$, where $k_+$ is
the heavy quark light cone residual momentum. $S(z)$ essentially
describes the probability for the $b$ quark to carry light cone momentum
fraction $z$ and allows for  a leakage past the partonic endpoint,
as can be seen  explicitly in the upper limit of $z$. 
A similar factorized expression holds
for the inclusive radiative decays near the endpoint,
\begin{eqnarray}
\frac1{\Gamma_0^\gamma}\frac{d\Gamma^\gamma}{d\hat s_0}
&=& \int_{1-\hat s_0}^{M_B/m_b} dz\; S(z)\, m_b^2 J[m_b^2(z-1+\hat s_0), \mu^2]
   \,H(m_b/\mu), \\
\Gamma_0^\gamma &=& \frac{G_F^2}{32\pi^4}|V_{ts}^*V_{tb}|^2\,
\alpha\;C_7^2\,m_b^5.
\end{eqnarray}

For the inclusive semi-leptonic decays, the integration over $x$ can
be done in the endpoint region and the resulting doubly differential
rate is\footnote{ Here we confirm explicitly that the dangerous region
$h \to 0$, where the energy of the hadronic jet vanishes and
factorization fails, is suppressed by the pre-factor and therefore not
important.}
\begin{equation}
\label{doubrate}
 \frac1{\Gamma_0}\frac{d^2\Gamma}{d\hat s_0\,dh} = 
 2 h^2 (3 - 2 h) \int_\xi^{M_B/m_b} dz\,S(z)\,m_b^2\,
  J[m_b^2\,h(z-\xi),\mu]\,H(m_b\,h/\mu) + {\cal O}(\hat s_0).
\end{equation}
This is where the factorization in the invariant mass spectrum is simpler
than in the electron energy spectrum. In the latter case, none of the 
integrals in the triply differential rate can be done trivially and one has 
to take an extra derivative with respect to $x$ 
to arrive at an expression similar to Eq.~(\ref{doubrate}).
An interesting consequence of factorization is that it connects the electron
energy spectrum in the region $x\to 1$ with the invariant mass spectrum,
in the region $\hat s_0 \to 0$,
\begin{equation}
\left.\frac{d\Gamma}{d\hat s_0}\right|_{\hat s_0 \to 1-x} =
 - \frac12 \; \frac{d}{dx}\, \frac{d\Gamma}{dx},
\end{equation}
which can be verified explicitly at the one loop level using the 
corresponding expressions in Ref.~\cite{FN}.

\section{The Perturbative Resummation}

At one loop level the differential rates are
\cite{FN,AG}
\begin{eqnarray}
\label{semione}
\frac1{\Gamma_0}\,\frac{d^2\Gamma}{d\hat s_0\,dh} &=&
 2 h^2\,(3 - 2h)\left[\delta(\hat s_0) + \frac{C_F \alpha_s}{4\pi}
    E_1(h, \hat s_0)\right] + 
   \frac{C_F \alpha_s}{4\pi} E_2(h, \hat s_0),\\
\label{radone}
\frac1{\Gamma_0^\gamma}\,\frac{d\Gamma^\gamma}{d\hat s_0} &=& 
\delta(\hat s_0)
 \left[ 1- \frac{C_F \alpha_s}{4\pi}\left(13 +\frac43\pi^2\right)\right]
  \nonumber\\
&& + \frac{C_F \alpha_s}{4\pi} \left[ 6 + 3\hat s_0
  -2\hat s_0^2 -2(2-\hat s_0)\log\hat s_0 - \left(\frac7{\hat s_0}
 + 4\, \frac{\log \hat s_0}{\hat s_0} \right)_+ \right],
\end{eqnarray}
where
\begin{eqnarray}
 E_1(h, \hat s_0) &=& - \delta(\hat s_0) \left[ 8 \log^2(h) 
 -10\log h + \frac{2\log h}{1-h} + 4\, {\rm Li}_2(1-h) + 5 
   +\frac43\pi^2 \right] \nonumber\\
 && -  4\left(\frac{\log \hat s_0}{\hat s_0} \right)_ +  +
  (8 \log h - 7)\left(\frac1{\hat s_0}\right)_+ +\frac1{\hat s_0}
  \left[ 8 \log\left(\frac{1+t}2\right) + 7(1-t)\right], \\
  E_2(h, \hat s_0) &=& \delta(\hat s_0)\, \frac{4 h^3 \log h}{1-h}
  - 4\left[2h (3-4h) -3 (1-2h)\hat s_0 -2 \hat s_0^2\right]
  \log\left(\frac{1+t}{1-t}\right) \nonumber\\
  && +\, 4\, h\, t \,(10 -15h +8 \hat s_0),
\end{eqnarray}
and $t = \sqrt{1-4\hat s_0/h^2}$. We also adopt the following definition
for the '+' distributions
\begin{equation}
\label{defplus}
\left(\frac{\log^n(\hat s_0)}{\hat s_0}\right)_+ = \lim_{\epsilon\to 0}
  \left[ \theta( \hat s_0 -\epsilon) \frac{\log^n(\hat s_0)}{\hat s_0} +
  \delta(\hat s_0)\, \frac{\log^{n+1}(\epsilon)}{n+1} \right].
\end{equation}
This definition is such that
\begin{equation}
\label{plus}
\int_0^\rho\, d\hat s_0\, F(\hat s_0)
  \left(\frac{\log^n(\hat s_0)}{\hat s_0}\right)_+ = 
  F(0)\,\frac{\log^{n+1}(\rho)}{n+1} \, + \, 
  \int_0^{\rho} \, d\hat s_0\, \left[ F(\hat s_0) - F(0) \right]
  \, \frac{\log^n(\hat s_0)}{\hat s_0}.
\end{equation}
Note that if the parameter $\rho$ becomes parametrically small, the
first term on the right hand side of Eq.~(\ref{plus}) will give large
logarithms thereby spoiling the systematics of the perturbative
expansions, while the second term must be regular as $\rho\to 0$.  To
 perform the resummation we go into the moment space where
the amplitudes factorize completely. In the case of inclusive
semi-leptonic decays, it is convenient to define a new variable
$\lambda= \hat s_0/h$ with kinematic range
\begin{eqnarray}
&& 0 \le h \le 1 \;; \quad 0 \le \lambda \le \frac{h}4, \nonumber\\
&& 1 \le h \le 2 \;; \quad 1-\frac1h \le \lambda \le \frac{h}4.
\end{eqnarray}
The region with $h \geq 1$ is populated with real gluon emissions only,
whereas the region with $h \leq 1$ has both real and virtual gluon
corrections. In the relevant  region $h\to 1$ and 
$\lambda \to 0$ the contributions from real and virtual gluon emissions 
combine to give terms which are '+' distributions, which upon
integrating
up to a cut, lead to the large logs we wish to resum.

To proceed, we take the $N$th moment with respect to
$\xi=1-\lambda$ in the large $N$ limit. In the region $\hat s_0 \sim 0$
and $z \sim \xi \sim 1$, one can replace $J[m_b^2\,h(z-\xi)]$ 
in Eq.~(\ref{doubrate}) with $J[m_b^2\,h(1-\xi/z)]$. This replacement
is permissible to the order we are working.
We then obtain
\begin{eqnarray}
\label{moment}
M_N &=& \int_0^1 d\lambda\;(1-\lambda)^{N-1}
   \frac1{\Gamma_0}\,\frac{d^2\Gamma}{d\lambda\,dh} \nonumber\\
&=& 2h^2(3-2h)\, S_N\, J_N(m_b^2\,h/\mu^2)\, H(m_b\,/\mu) + {\cal O}(1/N),\\
 J_N(m_b^2/\mu^2) &=& m_b^2 \int_0^1 dy\,y^{N-1} J[m_b^2(1-y),\mu], \\
 S_N &=& \int_0^{M_B/m_b} dz\; z^N S(z).
\end{eqnarray}
The soft moment $S_N$ further decomposes into a perturbative {\em soft}
piece, which accounts for soft gluon radiation and a non-perturbative piece
which incorporates bound state dynamics  and
serves as the boundary condition for the renormalization group
equation \cite{KS}.
\begin{eqnarray}
\label{deff}
S(z)&=& \int_z^{M_b/m_b} \frac{dy}{y} f[m_b(1-y)]\, \sigma(z/y), \\
S_N &=& f_N \,\sigma_N,
\end{eqnarray}
where $f(y)=\langle B(v)|\,\bar{b}_v\, \delta(y-iD_+)\, b_v\,|B(v)\rangle$
is the non-perturbative structure function defined in Ref.~\cite{N}.
A similar expression holds for the inclusive radiative decays 
$B\to X_s\gamma$
\begin{eqnarray}
\label{gamoment}
M_N^\gamma &=& \int_0^1 d\hat s_0\, (1-\hat s_0)^{N-1} \frac1{\Gamma_0^\gamma}
   \,\frac{d\Gamma^\gamma}{d\hat s_0} \nonumber\\
 &=& f_N\, \sigma_N\,J_N\, H^\gamma + {\cal O}(1/N),
\end{eqnarray}
Subsequently, we will ignore the non-perturbative structure function
$f(y)$ and
concentrate on the perturbative resummations.

It merits emphasizing that the large $N$ asymptotics of the moments
corresponds to the behavior of the spectra in the region $\hat s_0
\sim \lambda \sim 0$.  Taking the large $N$ limit also enables us to
extend the integration limit of $\lambda$ in Eq.~(\ref{moment}) up to
$1$, despite the fact that the kinematic range of $\lambda$ never goes
up to $1$. In this limit the contribution from the region $\lambda
\sim 1$ is power suppressed.

Comparing Eq.~(\ref{moment}) with the corresponding expression for the
electron energy spectrum in Ref.~\cite{KS}, one sees that the moments
$\sigma_N$ and $J_N$ are identical with those in the electron energy
spectrum, with change of variables $x\to 1-\lambda$ and $2-y_0\to h$.
A similar identification for the resummed radiative decays,
Eq.~(\ref{gamoment}), can be made with the change of variables $x\to
1-\hat s_0$. In moment space the soft and jet functions \cite{KT} have
been calculated to NLL order and are given by
\begin{eqnarray}
\sigma_N\,J_N &=& \exp [\,\log(N)\,g_1(\chi) + g_2(\chi)] \\
\sigma_N\,J_N^\gamma &=& \exp [\,\log(N)\,g_1(\chi) + g_2^\gamma(\chi)],
\end{eqnarray}
where $\chi = \alpha_s(m_b^2)\beta_0\log N$, and $g_1$ and $g_2$ are
given explicitly as \cite{AR}
\begin{eqnarray}
\label{g1}
g_1(\chi) &=& -\frac2{3\pi\beta_0\chi}
      [(1-2\chi)\log(1-2\chi) - 2(1-\chi)\log(1-\chi)], \\
\label{g2}
g_2(\chi) &=& g_2^\gamma(\chi) +  g_{sl}(\chi,h),\\
g_2^\gamma(\chi)&=&-\frac{k}{3\pi^2\beta_0^2}
            \left[\, 2\log(1-\chi)-\log(1-2\chi)\right]
   -\frac{2\beta_1}{3\pi\beta_0^3}\left[\log(1-2\chi)-2\log(1-\chi)
   \frac{}{}\right.\nonumber\\
&& + \left.\,\frac12\log^2(1-2\chi)-\log^2(1-\chi)\right]
   -\frac1{\pi\beta_0}\log(1-\chi)
   -\frac2{3\pi\beta_0}\log(1-2\chi) \nonumber\\
&& + \, \frac{4\gamma_E}{3\pi\beta_0}
    \left[\log(1-2\chi) - \log(1-\chi)\right],\\
g_{sl}(\chi,h) &=& \frac4{3\pi \beta_0} \log(h) \log(1-\chi).
\end{eqnarray} 
In the above, $\beta_0 = (11\,C_A - 2\,N_f)/(12\pi)$, $\beta_1 = 
(17\,C_A^2-5\,C_A\,N_f-3\,C_F\,N_f)/(24\pi^2)$,
$k = C_A\,(67/18-\pi^2/6) - 10\,T_R\,N_f/9$, and
$\gamma_E = 0.577216...$ is the Euler-Mascheroni constant. 
In our case, $C_A = 3$, $C_F = 4/3$, and $T_R = 1/2$.
To get back the physical spectra from the moment space, the inverse  
Mellin transform has to be evaluated at NLL accuracy as well. 
To this end, we apply
the identity derived in the Appendix of Ref.~\cite{LLR}
\begin{eqnarray}
\label{srinvmell}
&& \frac1{2 \pi i}\;\int_{C-i\infty}^{C+i\infty}\,dN\,x^{-N}\,
  e^{ \log (N)\,F_1[\alpha_s \log N ] + F_2[\alpha_s \log N ]} 
  \nonumber\\
&&\phantom{\frac1{2 \pi i}\;\int_{C-i\infty}^{C+i\infty}}=
  -x \frac{d}{dx}\;\left\{\theta(1-x) \frac{e^{l\:F_1(\alpha_s l) +
  F_2(\alpha_s l)}}
  {\Gamma\left[1 - F_1(\alpha_s l) - 
  \alpha_s l  \,F_1'(\alpha_s l)\right]}
    \times \left[1 + {\cal F}(\alpha_s ,l) \right] \right\},
\end{eqnarray}
where $l = -\log(-\log x) \approx -\log(1-x)$, and
\begin{equation}
{\cal F}(\alpha_s , l) = \sum_{k=1}^{\infty} \alpha_s^k \,
                      \sum_{j=0}^{k-1} f_{kj}\,l^j
\end{equation}
represents next-to-next-to-leading log contributions. 
Changing variables from $\lambda$ back to $\hat{s}_0/h$, we obtain
\begin{eqnarray}
\label{semispe}
\frac1{\Gamma_0}\,\frac{d^2\Gamma}{d\hat{s}_0 \,dh}&=&2 h^2 (3-2h)\,H(h)
 \nonumber \\ 
 && \times \frac{d}{d\hat{s}_0} \left\{\, 
 \theta\left(\frac{\hat{s}_0}h-\eta\right)\,
 \frac{ e^{l\:g_1(\alpha_s \beta_0\, l) + g_2(\alpha_s \beta_0\, l)}}
    {\Gamma\left[1-g_1(\alpha_s \beta_0\, l)-
   \alpha_s \beta_0\, l\, g_1'(\alpha_s \beta_0\, l)\right]}\right\}, \\
\label{radspe}
\frac1{\Gamma_0^\gamma}\,\frac{d\Gamma^\gamma}{d\hat s_0}&=& H^\gamma\,
   \frac{d}{d\hat s_0} \left\{ \,\theta(\hat s_0-\eta)\,
\frac{ e^{l\:g_1(\alpha_s \beta_0\, l^\gamma) + 
    g_2^\gamma(\alpha_s \beta_0\, l^\gamma)}}
    {\Gamma\left[1-g_1(\alpha_s \beta_0\, l^\gamma)-
 \alpha_s \beta_0\, l^\gamma\, g_1'(\alpha_s \beta_0\,l^\gamma)\right]}
 \right\},
\end{eqnarray}
where 
$l=-\log\left(-\log(1-\lambda)\right) \approx -\log(\hat{s}_0/h)$, and 
$l^\gamma=-\log\left(-\log(1-\hat s_0)\right) \approx
-\log(\hat s_0)$. The $\theta$-functions define the
differential rates in a distribution sense, as $\eta\to 0$, and turn
the singular terms into the '$+$' distributions, as can be seen 
explicitly by expanding in power series of $\log \hat s_0$ and using 
the definition Eq.~(\ref{defplus}).

The hard parts can be obtained through the one loop results
Eq.~(\ref{semione}) and Eq.~(\ref{radone})
\begin{eqnarray}
H(h) &=& 1-\frac{2\alpha_s}{3\pi}\left[ 4\log^2(h) - 5\log h + 
 \frac{3 \log h}{3-2h} +  2 {\rm Li}_2(1-h) + \frac52 +\frac{2\pi^2}3\right] \\
H^\gamma &=& 
1 - \frac{2\alpha_s}{3\pi}\left(\frac{13}2 + \frac{2\pi^2}3\right).
\end{eqnarray}
Eq.~(\ref{semispe}) and Eq.~(\ref{radspe}) 
reproduce the dominate contribution at one loop level
in the limit $\lambda\to 0$ and $\hat s_0\to 0$, respectively, and 
include the infinite set of terms of the form 
$\alpha_s^n \log^{n+1}(\hat{s}_0)$ and
$\alpha_s^n \log^{n}(\hat{s}_0)$ 
in the Sudakov exponent for both semi-leptonic and radiative decays.

\section{The Integrated Cut Invariant Mass Spectrum}

As previously mentioned,  it has been proposed that we 
 measure the modulus of the CKM matrix element
$V_{ub}$ from inclusive decays by making a cut on the hadronic
invariant
mass below $M_D^2$. This cut  eliminates the overwhelming
background
from  bottom to charm transitions. While it is the hadronic invariant
mass which is of interest, we shall first consider the cut partonic
invariant mass, as it will be relevant to our conclusions.

The use of an upper cut $c_0$ on the partonic invariant mass 
introduces large logs of the form $\alpha_s \log^2(c_0)$. 
As $c_0$ approaches zero, the logs become parametrically
large and need to be resummed. Using the resummation formulas
in the previous section, it is simple to generate an expression for
the resummed cut rate, since our expression can be written as
a total derivative with respect to $\hat{s}_0$. The cut rate may be written
as

\begin{equation}
\label{partcut}
\frac1{\Gamma_0}\,\Gamma(c_0)=
\int_0^1~ dh\;2  h^2(3-2h) \left[\xi\left(\frac{c_0}{h}\right)-
  \xi_{\alpha_s}\left(\frac{c_0}{h}\right)\right] 
  +\int_0^{c_0}
d\hat{s}_0\,\int^{1+\hat{s}_0}_{2\sqrt{\hat{s}_0}}
dh \;\gamma_{\alpha_s}(h,\hat{s}_0),
\end{equation}
where 
\begin{equation}
\xi(c_0/h)=\left.\frac{ e^{l\:g_1(\alpha_s \beta_0\, l) + 
g_2(\alpha_s \beta_0\, l)}}
    {\Gamma\left[1-g_1(\alpha_s \beta_0\, l)-
   \alpha_s \beta_0\, l\, g_1'(\alpha_s \beta_0\, l)\right]}
\right|_{l=-\log(c_0/h)},
\end{equation}
and $\gamma_{\alpha_s}(h,\hat{s}_0)$ is simply the one loop rate defined
in Eq.~(\ref{semione}). $\xi_{\alpha_s}$ stands for terms up to
${\cal O}(\alpha_s)$ when expanding $\xi$ in power series of $\alpha_s$.
We subtracted it from $\xi$ in order to ensure that we correctly reproduce
the one loop result at order $\alpha_s$. From Eq.~(\ref{partcut}) one can
see that the large logs arise when $\hat{s}_0$ approaches zero, which is not
only the lower kinematic limit for $\hat{s}_0$, but also the phase space
boundary for virtual and real gluon emissions.

The perturbative expansion has been reorganized into an expansion in
the exponent. The systematics of this expansion have been discussed at
length in \cite{LLR,LR}. Here we just recall that we may test the
convergence of the reorganized expansion by comparing the NLL resummed
result with the leading-logarithm (LL) resummed result. Fig.~2 shows
the cut rate as a function of $c_0$. In this figure we show the one
loop cut rate as well as the resummed cut rate with and without NLL
corrections. We see that for $0.1 < c_0 < 0.2$ the resummation becomes
necessary, while for $c_0<0.1$ the NLL dominates the LL, so that we
can no longer trust our results. Indeed, this breakdown occurs well
before we reach the Landau pole at $\hat{s}_p=e^{-1/(2\alpha_s
\beta_0)} \approx 0.028$.

\begin{figure}[t]
\centerline{\epsfysize=11truecm  \epsfbox{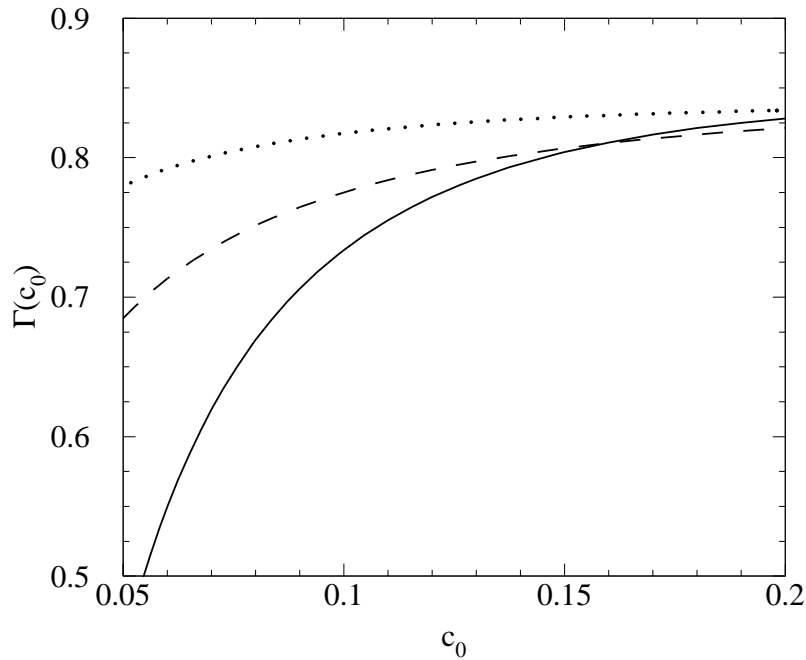} }
\tighten{
\caption[]{\it The rate as a function of the partonic cut. The dotted
line is the one loop result, the dashed line is the LL result, while
the solid line is the NLL log result. The difference between the
one loop and resummed results at large $c_0$ is due to two loop
corrections introduced in the resummation. We use
$\alpha_s(m_b)=0.21$.}}
\end{figure}

We now expand this result to pick out the leading and
next to leading infrared log contribution to the two loop
differential rate. This contribution is given by
\begin{equation}
\label{BLM}
\frac1{\Gamma_0}
\left.\frac{d\Gamma}{d\hat{s}_0dh}\right|_{{\cal O}(\alpha_s^2)}
= 2 h^2(3-2h)\,
\frac{\alpha_s^2}{3 \pi^2}\, \left[ \frac{8}{3}
\left(\frac{\log^3(\hat{s}_0)}{\hat{s}_0}\right)_+ + (14 - 6 \pi
\beta_0) \left(\frac{\log^2(\hat{s}_0)}{\hat{s}_0}\right)_+ \right]
\end{equation}
As expected, we see that the most singular contribution at 
${\cal O}(\alpha_s^2)$ doesn't have any terms proportional
to $\beta_0$. It may be the case that, in this particular region
of phase space the infrared logs terms may dominate over the
BLM terms.
Such a conclusion
was reached in \cite{LLR,LR} for the two loop contribution to lepton
and photon spectra in  semi-leptonic and radiative decays, respectively. 
Naturally
, this
does not preclude the possibility that there exist a cancellation 
with other uncalculated terms, such that the $\alpha_s^2 \beta_0$ 
still dominate.

Let us now consider the physical case, where we are interested in
placing an upper cut on the {\it hadronic} invariant mass. 
The hadronic invariant mass may be written as

\begin{eqnarray}
\label{shcut}
\hat s_H &=& \frac{s_H}{m_b^2} = \hat s_0 + \epsilon h + \epsilon^2, \\
\epsilon &=& \frac{\bar\Lambda}{m_b},
\end{eqnarray}
where $\bar{\Lambda}$ is the mass difference $M_B - m_b$ in the
infinite $b$ quark mass limit, which is a measure of
binding energy for the $b$ quark inside the $B$ meson. 
Thus, given a cut on $\hat{s}_H$,  $c$, we may translate 
this into an $h$ dependent cut on $\hat{s}_0$.
After changing the order of integration we find that
 the  cut rate may be written as 
\begin{equation}
\label{upper}
\frac1{\Gamma_0} \Gamma(c)=
\int^{\frac{c-\epsilon(1+\epsilon)}{1+\epsilon}}_0 d\hat{s}_0
\int_{2\sqrt{\hat{s}_0}}^{1+\hat{s}_0}dh\left(\frac{d\Gamma}{d\hat{s}_0dh}\right)  +
\int_{\frac{c-\epsilon(1+\epsilon)}{1+\epsilon}}^{(\sqrt{c}-
\epsilon)^2} d\hat{s}_0 
\int^{\frac{c-\hat{s}_0-\epsilon^2}
{\epsilon}}_{2\sqrt{\hat{s}_0}}dh
\left(\frac{d\Gamma}{d\hat{s}_0dh}\right),
\end{equation}
if  $c>\epsilon(1+\epsilon)$,
whereas if  $c<\epsilon(1+\epsilon)$ then  
\begin{equation}
\frac1{\Gamma_0} \Gamma(c)=
\int^{(\sqrt{c}-\epsilon)^2}_0 d\hat{s}_0
\int_{2\sqrt{\hat{s}_0}}^{\frac{c-\hat{s}_0-\epsilon^2}
{\epsilon}}dh \left(\frac{d\Gamma}{d\hat{s}_0dh}\right).
\end{equation}

This situation is different from the partonic invariant mass
spectrum. In the present case, the lower kinematic limit of
$\hat{s}_H$ is $\epsilon^2$, while the phase space boundary for
virtual and real gluon emissions has been moved to
$\rho_{\epsilon}=\epsilon(1+\epsilon)$, above which only real gluon
emissions contribute.\footnote{A more general discussion for this kind
of phenomenon in other observables can be found in
Ref.~\cite{shoulder}.}  Hence there could potentially be two kinds of
parametrically large logarithms $\log(c-\epsilon^2)$ and
$\log(c-\rho_{\epsilon})$ appearing. However, as can be seen from
Eq.~(\ref{shcut}), $\hat{s}_H \to \epsilon^2$ corresponds to vanishing
$\hat{s}_0$ and $h$, which is the dangerous region where the infra-red
factorization fails and is kinematically suppressed by the tree level
rate. More explicitly, $\log(c-\epsilon^2)$ will be killed by the
pre-factor $2(c-\epsilon^2)^2$ so that the rate vanishes as $c \to
\epsilon^2$. The only important parametrically large logarithms are
those resulting from the incomplete cancellation between virtual and
real corrections at the phase space boundary $\rho_{\epsilon}$. From
this argument it would seem that we only need to resum logs of the
form $\log(c-\rho_\epsilon)$. However, in this particular case, the
distance from $\rho_\epsilon$ to $\epsilon^2$ is only $\epsilon$,
which is expected to be a numerically small quantity at around
$0.08$. If the experimental cut $c$ lies below $\rho_\epsilon$, the
partially integrated rate could still be sensitive to
$\log(c-\epsilon^2)$ due to the smallness of $\epsilon$.  We thus
chose to resum all logs of the form $\log(c-\epsilon^2)$ as well.
Fortunately, since we have the resummed rate at the doubly
differential level, this is not a problem.  The resummed rate with
hadronic mass cut $c>\rho_\epsilon$ is given by
\begin{eqnarray}
\label{hadcutI}
\frac1{\Gamma_0} \Gamma(c)_> =& &\int_0^1~ dh \,2 h^2(3-2h)\,
\left\{\xi\left[\frac{c-\rho_\epsilon}{h(1+\epsilon)}\right]-
  \xi_{\alpha_s}\left[\frac{c-\rho_\epsilon}{h(1+\epsilon)}\right]
\right\} \nonumber \\
&+& \int_{\frac{c-\rho_\epsilon}{1+\epsilon}}^
{(\sqrt{c} - \epsilon)^2}d\hat{s}_0
\int_{2\sqrt{\hat{s}_0}}^{\frac{c-\hat{s}_0-\epsilon^2}{\epsilon}}dh
 \, 2 h^2(3-2h)\,
\left[\xi^\prime(\hat{s}_0/h)-\xi^\prime_{\alpha_s}(\hat{s}_0/h)\right]\nonumber \\
&+&
\int_0^{\frac{c-\rho_\epsilon}{1+\epsilon}}
d\hat{s}_0\int_{2\sqrt{\hat{s}_0}}^{1+\hat{s}_0}dh\; 
\gamma_{\alpha_s}(h,\hat{s}_0) +
\int_{\frac{c-\rho_\epsilon}{1+\epsilon}}^
{(\sqrt{c} - \epsilon)^2}d\hat{s}_0
\int_{2\sqrt{\hat{s}_0}}^{\frac{c-\hat{s}_0-\epsilon^2}{\epsilon}}dh
\; \gamma_{\alpha_s}(h,\hat{s}_0),
\end{eqnarray}
where 
\begin{equation}
\xi^\prime(\hat{s}_0/h)=\frac{d}{d\hat{s}_0}\xi(\hat{s}_0/h).
\end{equation}
While the resummed rate with hadronic mass cut $c<\rho_\epsilon$
is given by 
\begin{eqnarray}
\label{hadcutII}
\frac1{\Gamma_0} \Gamma(c)_<=&& 
\int_0^\frac{c-\epsilon^2}{\epsilon} dh\;2 h^2(3-2h)\,
\left\{\xi\left[\frac{(\sqrt{c}-\epsilon)^2}{h}\right] -
 \xi_{\alpha_s}\left[\frac{(\sqrt{c}-\epsilon)^2}{h}\right]\right\} 
 \nonumber\\
&+&\int_0^{\left(\sqrt{c}-\epsilon\right)^2}
d\hat{s}_0 \int_{2\sqrt{\hat{s}_0}}^{1+\hat{s}_0}
\,\gamma_{\alpha_s}(h,\hat{s}_0).
\end{eqnarray} 
Analytic expressions for the partially integrated rate at one 
loop level, last lines in Eq.~(\ref{hadcutI}) and Eq.~(\ref{hadcutII}),
can be found in Ref.~\cite{FN}.

\begin{figure}[t]
\centerline{\epsfysize=11truecm  \epsfbox{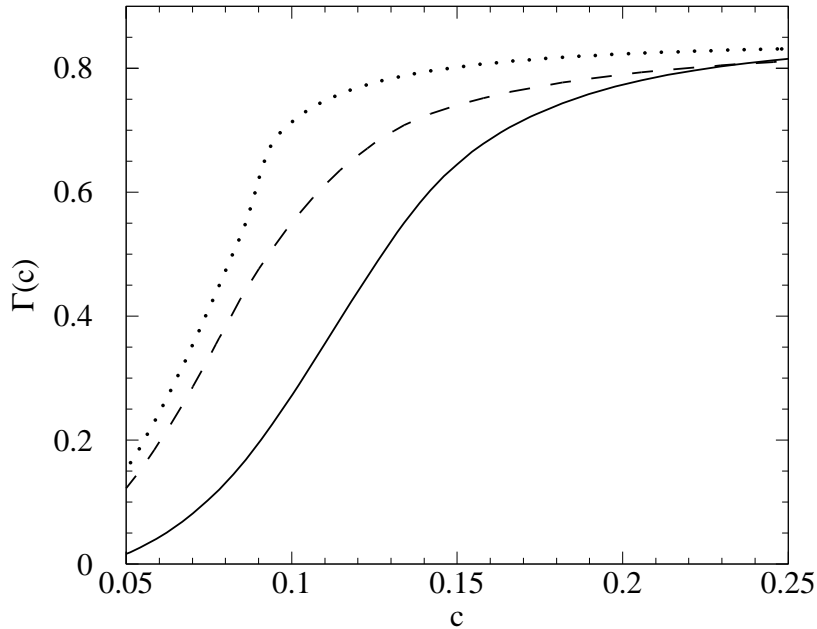} }
\tighten{
\caption[]{\it The rate with a hadronic cut. The dotted line is the
one loop result, the dashed line is the LL resummed result, while the
solid line is the NLL resummed result.  For $\bar{\Lambda}= 0.39\,{\rm
GeV}$, we have $\rho_\epsilon=0.0861$.  In the region above
$\rho_\epsilon$, we run into Landau pole at $c_p = 0.1159$ and
interpolate in the small region between $\rho_\epsilon$ and $c_p$.}}
\end{figure}
\section{Results and Conclusions}
In Fig.~3 we show the one loop, LL resummed (including $g_1$ only) and
NLL resummed (including $g_1$ and $g_2$) results for
$\bar{\Lambda}=0.39\,\mbox{GeV}$.  We see the for $c\leq 0.18$, the
next to leading order approximation breaks down, as the next to
leading order piece becomes just as large as the leading order piece.
Notice that when $c\simeq 0.18$, the effective cut on the partonic
invariant mass is $c_0 \simeq 0.09$, which from Fig.~2 we see is
consistent with the breakdown of the resummed expansion.  This is
contrast with the result that an energy cut of $2.1 {\rm\ GeV}$, on
the rate for $B\rightarrow X_s \gamma$ does not necessitate
resummation, as the argument of the logs in the case of the radiative
decay is about $0.12$.

It is clear from Fig.~3 that resummation shifts the whole spectrum
toward the high invariant mass region such that the number of events
which lie below the cut $c$ is decreased.  This occurs because the
high invariant mass region with $\hat{s}_H > \rho_\epsilon$ is
populated with real gluon emissions only.

\begin{figure}[t]
\centerline{\epsfysize=11truecm  \epsfbox{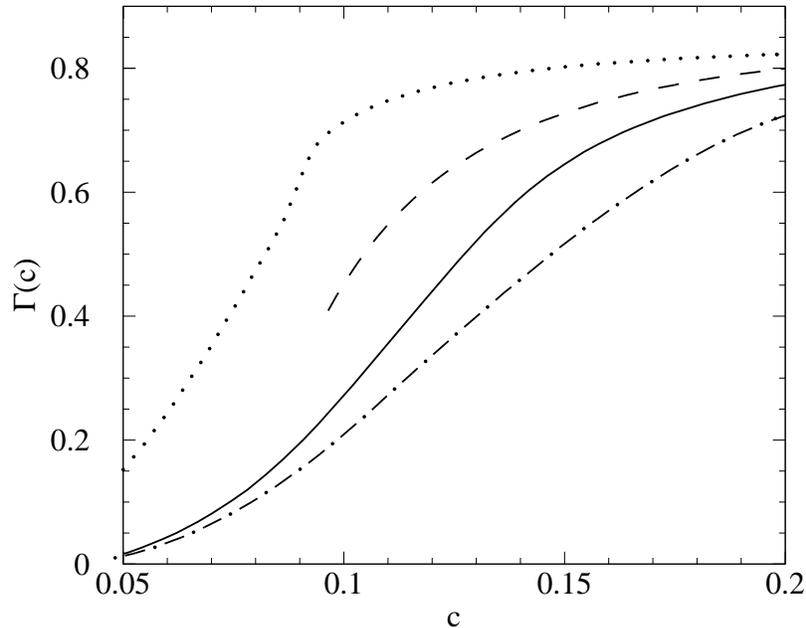} }
\tighten{
\caption[]{\it The cut rate for several different values of
$\bar{\Lambda}$. The dashed line is for $\bar{\Lambda} = 0.28\,{\rm
GeV}$, the dot-dashed line is for $\bar{\Lambda} = 0.50\,{\rm GeV}$,
while the solid line is for $\bar{\Lambda} = 0.39\,{\rm GeV}$. The
dotted line is the one loop result for $\bar{\Lambda} = 0.39\,{\rm
GeV}$.}}
\end{figure}

In Fig.~4 we show the cut rate for several different values of
$\bar{\Lambda}$. It is clear that the cut rate can be very sensitive
to the value of the unphysical parameter $\bar{\Lambda}$. This occurs
because the argument of the large logs is now $c-\rho_\epsilon$. The
parameter $\bar{\Lambda}$ is well defined at a given fixed order in
perturbation theory and the values chosen in Fig.~3 are the mean value
and one sigma values extracted in Ref.~\cite{GKLW} at one loop. Soon
the errors in the extraction of $\bar{\Lambda}$ will become less
significant. In this paper we will not delve into the phenomenology of
the extraction of $V_{ub}$, as the calculation we have discussed here
has not included the important non-perturbative corrections coming
from the Fermi motion. These corrections are parameterized in terms of
a well defined structure function, $f(y)$ defined in Eq.~(\ref{deff}),
and as discussed in \cite{FN}, can be very large. We will reexamine
the issue of the structure function in a future publication, where
following \cite{LLR}, we will eliminate the structure function from
the cut rate prediction by utilizing the data from the end-point of
radiative $B$ decays, which in turn encodes all the information
contained in the structure function. 

From Fig.~4 we see 
that, for large values of $\bar{\Lambda}$, the resummation is not under
control for phenomenologically interesting cuts and thus it may  not be  
possible 
to extract $V_{ub}$ in this way. However, when modding out by the soft 
function using the  $b\to s\gamma$ rate, there are cancellations 
which lead to a perturbative series which is better behaved. 
This is indeed what happened in the
case for the electron energy spectrum\cite{LLR}. Thus, we refrain from
drawing any conclusions regarding the viability of extracting 
$V_{ub}$ from the invariant mass spectrum at this time.

The purpose of this paper was to determine the effect of threshold
resummation on the rate for semi-leptonic $B$ decays with a cut on the
hadronic invariant mass. We first showed how the rate factorizes when
written in terms of hadronic variables, generalizing the results of
\cite{KS}. Using this factorization, we resummed the cut rate at next
to leading order in the infrared logs and found that, for cuts of
interest, the resummation is crucial, and that for $c<0.18$, even the
next to leading order resummed rate is no longer reliable.  However,
this breakdown point depends on the value of $\bar{\Lambda}$, and
becomes smaller as $\bar{\Lambda}$ is decreased. A more
phenomenological analysis, including the effects of the structure
function responsible for the Fermi motion, is forthcoming.

\acknowledgments 
This work was supported in part by the Department of Energy under
grant number DOE-ER-40682-143. I.\,L. would like to thank the hospitality
of the National Center for Theoretical Sciences at National Tsing Hua
University in Taiwan where part of this work was completed.

\tighten


\end{document}